\documentclass[5p]{elsarticle}

\usepackage[T1]{fontenc}
\usepackage[ansinew]{inputenc}
\usepackage{booktabs}
\usepackage{lmodern} 

\usepackage{microtype}

\usepackage{graphicx}
\usepackage{upgreek}
\usepackage{amsmath,amssymb}
\usepackage[breaklinks]{hyperref}

\newcommand{\ghz}{\ensuremath{\, \mathrm{GHz}}}
\newcommand{\mhz}{\ensuremath{\, \mathrm{MHz}}}
\newcommand{\khz}{\ensuremath{\, \mathrm{kHz}}}
\newcommand{\dbm}{\ensuremath{\, \mathrm{dBm}}}
\newcommand{\Oe}{\ensuremath{\, \mathrm{Oe}}}
\begin{document}

\begin{frontmatter}

\title{Onset of spin wave time-domain fractals in a dynamic artificial crystal}


\author[1]{Alistair Inglis\corref{cor1}}
\ead{alistair.inglis@physics.ox.ac.uk}

\author[2]{John F. Gregg}
\ead{john.gregg@magd.ox.ac.uk}

\cortext[cor1]{Corresponding author}
\address{Department of Physics, University of Oxford, Clarendon Laboratory, Parks Road, OX1 3PU}



\date{\today}

\begin{abstract}
We report on the first observation of an exact fractal pattern in the time-domain arising spontaneously from a dynamic artificial crystal (DAC). The all-magnon process occurs in a passive, unlithographed magnetic waveguide. The DAC was created by a standing spin wave of frequency $f_{\mathrm{DAC}}$ in a region of nonlinear waveguide, resulting in a spatio-temporally periodic potential. The interaction of travelling spin waves of frequency $f'$ with the DAC resulted in a series of new modes appearing in a comb with intervals $\Delta f^{(0)} = \mid f_{\mathrm{DAC}}-f'\mid$. As $H$ was increased a 1st pre-fractal pattern was observed with frequency interval $\Delta f^{(1)} =\Delta f^{(0)}/2$. Finally, the onset of a 2nd pre-fractal was observed with frequency interval $\Delta f^{(2)} =\Delta f^{(1)}/2 = \Delta f^{(0)}/4$. The magnetic field dependence of the nonlinear signals effect matched that of the DAC. The fractal-like behavior was demonstrated for different values of $f'$. 
\end{abstract}





\end{frontmatter}

\section{Introduction}
The study of fractals began with Benoit Mandelbrot in the 1960s and has been of scientific interest ever since. The phenomenon of fractal behavior, or self-similarity, exists in a remarkable range of observable physical systems. Indeed, fractal analysis may be applied to such varied systems including crystal growth, diffusion fronts, lung structure, moon crater distribution, stock market fluctuations and authenticating priceless artwork \cite{Addison1997,Mandelbrot1984,Taylor2007}. The ubiquity of fractals points to their fundamentality, and the simplicity with which they may be described is testament to their beauty.

Fractals may be subdivided into two main categories: exact and statistical. An exact fractal is an object that exhibits a perfect symmetry of scale: it replicates the structure of the whole upon varying degrees of magnification. The application of this general condition can  generate some rather striking images; famous examples include the Sierpinski gasket and the Mandelbrot set.

A statistical fractal on the other hand, is an object in which the self-similarity is only present in the statistical properties of the system. Perhaps intuitively, it is these fractals that are vastly more common in Nature as they allow for random deviations from the exact regime, which in contrast is relatively rare.

Despite the rarity however, exact fractals have been observed in nonlinear dynamical systems \cite{Segev2012}. Space-domain soliton fractals were first demonstrated numerically using nonlinear optics \cite{Soljacic2000,Sears2000} with experiments demonstating similar results \cite{Erkintalo2012}, while time-domain soliton fractals were experimentally realised within a magnetic medium\cite{Wu2006}. Until recently \cite{Richardson2018}, all observed exact time-domain fractals were exclusively soliton in nature.  Richardson \textit{et al.} have demonstrated however, that spontaneous continuous wave (CW) time-domain fractals can emerge in magnonic waveguides possessing highly nonlinear dispersion, as in a quasi-1D magnonic crystal. In their experiment, the artificial crystal had periodic etched grooves which created a rejection band for magnons of wavelengths associated with the periodicity of the grooves. The altered transmission band corresponded to a sharp kink in the dispersion curve \cite{Ordonez-Romero2016}, and consequently a large spike in the dispersion coefficient resulting in the enhanced nonlinearity which facilitated the formation of fractals when driven at high power. In this case, the fractal behavior originated from self modulational instability (SMI) \cite{Zakharov2009} where the time-domain side bands represented the amplitude modulation. Significantly, these fractals appeared spontaneously out of the passive waveguide element, rather than being forced into existence \cite{Richardson2018}.

\begin{figure}[htbp]
	\centering
		\includegraphics[width=0.45\textwidth]{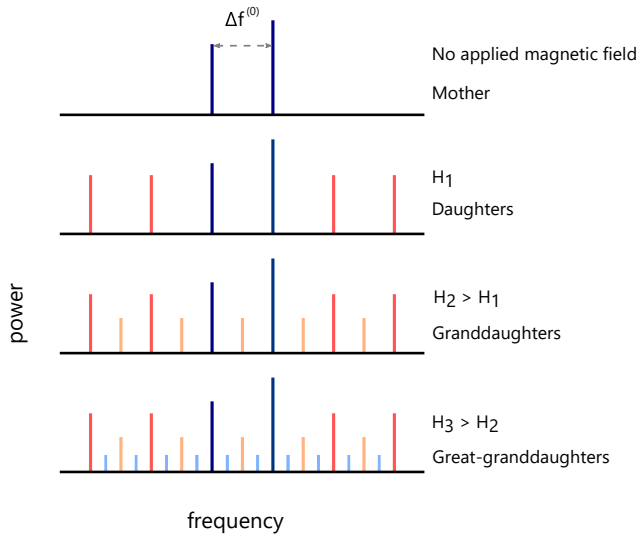}
	\caption{Sketch of the development of observed fractal pattern. Two inputs form the mother mode, comprising two frequencies detuned by $\Delta f^{(0)}$ which are present at zero field. At certain field $H_{1}$ a comb of peaks separated by $\Delta f^{(0)}$, forming the daughter modes. At $H_{2}$, a new comb of interval $\Delta f^{(1)} =\Delta f^{(0)}/2$ appears. Then, at $H_{3}$, a new comb of interval $\Delta f^{(2)} =\Delta f^{(1)}/2$ is formed. }	
	\label{fig:fractalsketch}
\end{figure}

In our experiment, the time domain fractals, shown in Fig. \ref{fig:fractalsketch}, manifest as follows: (1) there are two initial input signals which form the initiator, also known as the mother.  (2) A comb of sidebands is excited which, in fractal parlance, represents the generator, or daughters. (3) At the mid-points between consecutive comb peaks, further granddaughter peaks appear. (4) Half-way between daughter and granddaughter peaks, great-granddaughter peaks were observed.

At this stage, a note on terminology should be made for the uninitiated - the granddaughters and great-granddaughters may be referred to as 1st and 2nd pre-fractals, respectively. The term pre-fractal simply refers to an exact fractal that is not infinitely scalable (as is necessarily the case for all exact fractals in the physical world). Despite this slightly imprecise language, we refer to the observed object as a fractal proper, as is convention \cite{Wu2006,Richardson2018,Sears2000,Soljacic2000}. 

The time-domain fractal pattern may also be understood as successive rounds of amplitude modulation, arising as a consequence of modulational instability (MI). For MI to occur, the Lighthill criterion must be satisfied \cite{Zakharov2009}, namely

\begin{equation}
\bigg(\frac{\partial^{2}\omega}{\partial k^{2}}\bigg)\bigg(\frac{\partial\omega}{\partial |a|^2}\bigg) < 0
	\label{lighthill}
\end{equation}
where $k$ is the magnon wavenumber, and $a$ is the magnon amplitude. 

In this work, we report on the first observation of time-domain fractals using a dynamic artificial crystal (DAC) in a simple, isotropic thin film, free from patterning and lithography. This demonstration contrasts with previous reports of fractals in magnetic media in that it does not utilize a permanent artificial crystal. Furthermore, the frequency comb that defines the fractal pattern is demonstrated to be tunable, existing over a range of frequencies. While the precise mechanism for the presence of fractals is unclear, we posit an explanation not inconsistent with previous reports.

\section{Experimental Method and set-up}
%
%

\begin{figure}[htbp]
	\centering
		\includegraphics[width=0.45\textwidth]{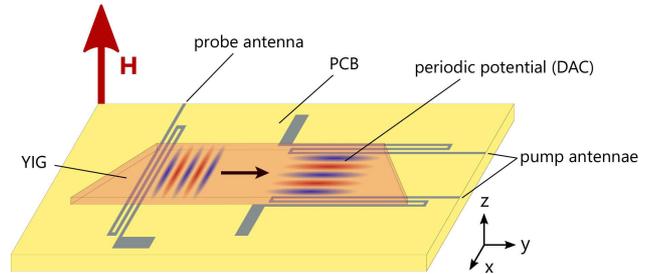}
	\caption{Experimental configuration: YIG film with 45$^{\circ}$ edges placed on PCB with antennae. Counter-propagating pumps excite a standing spin wave creating a spatio-temporally periodic potential due to intensity dependent refractive index. The probe antenna transmits excitation magnons and detects spin waves interacting with the DAC region.}
	
	\label{fig:yigsetup}
\end{figure}
\subsection{General configuration}
A schematic of the experiment is shown in Fig. \ref{fig:yigsetup}. For a magnon waveguide we used a thin film of yttrium iron garnet(YIG) that was $18 \, \mathrm{mm}$ long, $2.1 \, \mathrm{mm}$ wide, and $7.8 \,\upmu$m thick, backed on a substrate of gadolinium gallium garnet (GGG), with edges cut at $45^{\circ} $. This was placed YIG-side-down on a printed circuit board (PCB) whereby two parallel pump antennae were used to excite a dynamic artificial crystal (DAC) \cite{Karenowska2012,Chumak2010a} across the region of waveguide subtended by the pump antennae. A probe antenna was placed approximately $3\, \mathrm{mm}$ away, oriented perpendicular to the former antennae. 

\subsection{Antenna design and charactersiation}

\begin{figure}[htbp]
	\centering
		\includegraphics[width=0.49\textwidth]{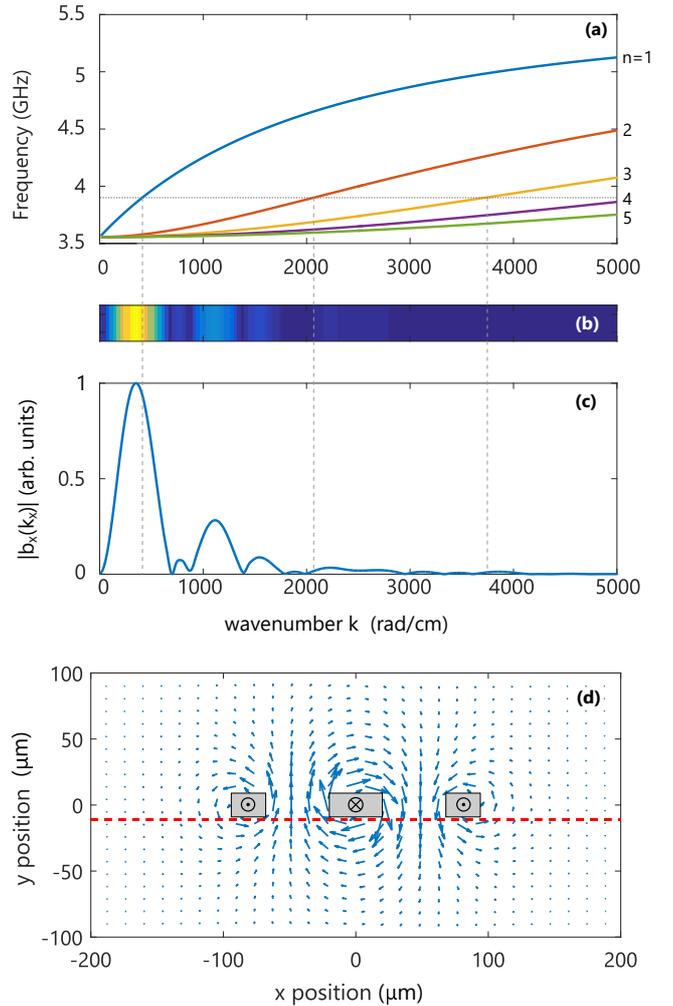}
	\caption{(a) FVMSW dispersion curves for the first five thickness modes calculated for a YIG film of thickness $d=7.8\, \upmu$m $M_{\mathrm{sat}} = 140\,\mathrm{kAm}^{-1}$, and an applied magnetic field of $H=3050 \Oe$. (b)-(c) Map of the relative amplitude of magnetic field induced by antenna as a function of spatial frequency. (d) Field distribution calculated for the fabricated meander structure antennae. Red dashed line shows field plane for which Fourier transform in (c) was performed. }	
	\label{fig:FVmodesAndAntenna}
\end{figure}

The microwave antennae were used to inductively excite a type of dipolar magnon known as forward volume magnetostatic spin waves (FVMSWs) which have an isotropic dispersion (in contrast to the highly anisotropic dispersion of backward volume magnetostatic spin waves and surface spin waves  \cite{Schneider2010,Demidov2009c}). Magnons with an isotropic dispersion were necessary due to the mutually orthogonal arrangement of the pump and probe antennae. The dispersion curves of the first five FVMSW thickness modes were calculated using the result of Kalinikos and Slavin \cite{Kalinikos1986a} for the YIG film described above, with saturation magnetization $M_{\mathrm{sat}} = 140\,\mathrm{kAm}^{-1}$, and an applied magnetic field of $H=3050 \Oe$. Presented in Fig. \eqref{fig:FVmodesAndAntenna}(a), it is of note that the $n=1$ mode has a dispersion such that the first term in Eqn.(\ref{lighthill}) is always negative, while the second term is always positive\cite{Boardman1995}, thus satisfying the Lighthill criterion and allowing the possibility of MI. 

All antennae were meander structures designed to suppress the $k=0$ spin wave mode. There were three main reasons for this: (1) spin waves with larger wavenumbers are less prone to diffraction, improving the confinement of the standing wave that defines the DAC to the region of waveguide subtended by the antennae. (2) The $k=0$ mode corresponds to FMR which will not create a DAC. (3) At larger wavenumbers the antenna coupling to higher magnon thickness modes for a given frequency becomes negligible, allowing us to consider only the $n=1$ thickness mode. 

An identical meander pattern was used for all the antennae. This consisted of three legs of length $4\,$mm and individual legs $40\,\upmu$m wide, with a spacing of $50\,\upmu$m between legs. Using a Biot-Savart method, the cross-sectional  field distribution of a fabricated antenna was calculated, and is shown in Fig. \ref{fig:FVmodesAndAntenna} (d). From this, information  about the spatial frequencies of the field induced by the antenna was obtained. This is presented in Fig. \ref{fig:FVmodesAndAntenna} (c), where $|b_{x}(k_{x})|$ represents the Fourier transform of the $x$-component of the magnetic field, $b_{x}(x)$, along the $x$-direction, for the magnetic field $2\,\upmu$m into the film, denoted by the dashed red line in (d). It is demonstrated that for our antenna design, $|b_{x}(k_{x})|$ approaches zero in the limit where $k$ goes to zero. Since the excitation efficiency of spin wave modes scales linearly \cite{Demidov2009b} with $|b_{x}(k_{x})|$, we can assume contributions from the FMR mode may be neglected. Furthermore, the effects of  any thickness modes above the $n=1$ mode were neglected for two reasons. The first may be seen from (a) and (c), where we note that for a typical excitation frequency used in experiment ($3.93\ghz$) the difference in $|b_{x}(k_{x})|$ between the $n=1$ and $n=3$ modes was over 2 orders of magnitude. Secondly, the spin wave excitation efficiency is also proportional to the convolution of the magnetic field and the amplitude profile of the thickness mode \cite{Demidov2009b}, which is zero for even modes (by symmetry) and scales as $1/n$ for odd modes \cite{Phillips1966}, assuming that $\lambda_{0} \gg d$, where $\lambda_{0}$ is the free space wavelength, and $d$ is the film thickness.

The antenna efficiency with respect to frequency was investigated using a Rhode $\&$ Schwarz ZVL Network Analyzer. The $S_{11}$ values for each antenna were found to have local minima at approximately $3.93\ghz$, as shown in Fig. \ref{fig:FVmodesAndAntenna}. Since the minimum in $S_{11}$ corresponds to a minimum of reflected power, frequencies in this region were used for the experiments described below. 

\begin{figure}[htbp]
	\centering
		\includegraphics[width=0.49\textwidth]{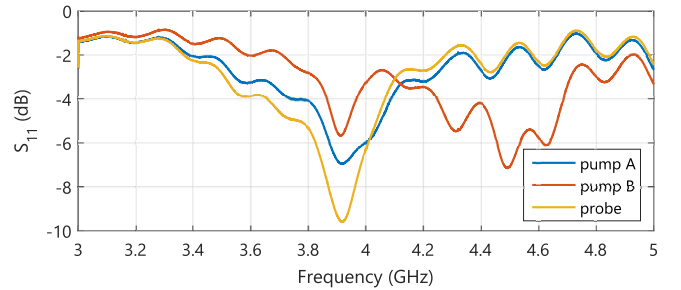}
	\caption{Measured $S_{11}$ values of the pump and probe antennae.}	
	\label{fig:S11_Antennae}
\end{figure}

\subsection{Characterization of DAC}

To create the DAC, we introduced a spatio-temporally periodic potential by exciting a standing spin wave across the width of the waveguide \cite{Bloom1977a}. Due to the well-documented nonlinearity of YIG \cite{Temiryazev1989,Nikitov1995,Slavin2002,Drozdovskii2016,Boardman1988}, the local refractive index of the film depends on the amplitude of the magnons in the given region \cite{Boardman1995}. The periodic change in refractive index resulting from the standing wave may be considered as an effective-grating that, as such, alters the transmission efficiency of magnons with wavelength relating to that of the effective-grating spacing. In turn, this significantly modifies the dispersion at that given wavenumber, facilitating fractal behavior \cite{Richardson2018}.

Spin wave reflections due to the waveguide edges and pump antennae cause the pumped region to behave like a resonant cavity, wherein a standing wave will only be supported for specific wavenumbers. The standing wave comprising the DAC will be active when the applied magnetic field strength shifts the dispersion such that the pump magnons have a wavenumber, $k_{\mathrm{DAC}}$, that satisfies
\begin{equation}
k_{\mathrm{DAC}}(\omega,H) = \frac{\pi N}{w},
\end{equation}  
where $w$ is the width of the waveguide and $N$ is an integer. If this condition is not met, side-wall reflections will destroy the standing wave and hence the DAC. 

\begin{figure}[htbp]
	\centering
		\includegraphics[width=0.45\textwidth]{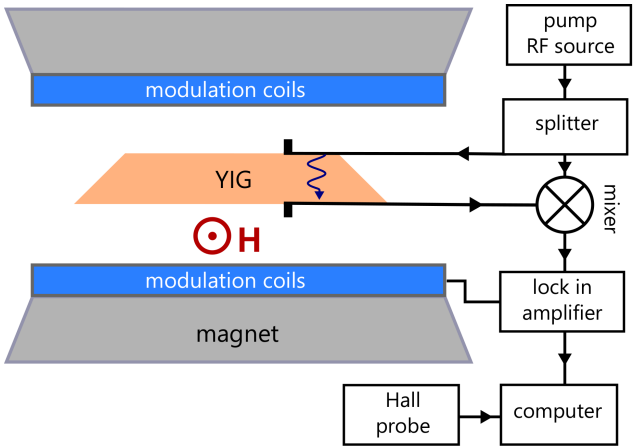}
	\caption{Experimental set-up for characterizing the standing spin wave and DAC field dependence using a homodyne technique.}	
	\label{fig:blockdiagramschematicHomodyneV1}
\end{figure}

\begin{figure}[htbp]
	\centering
		\includegraphics[width=0.49\textwidth]{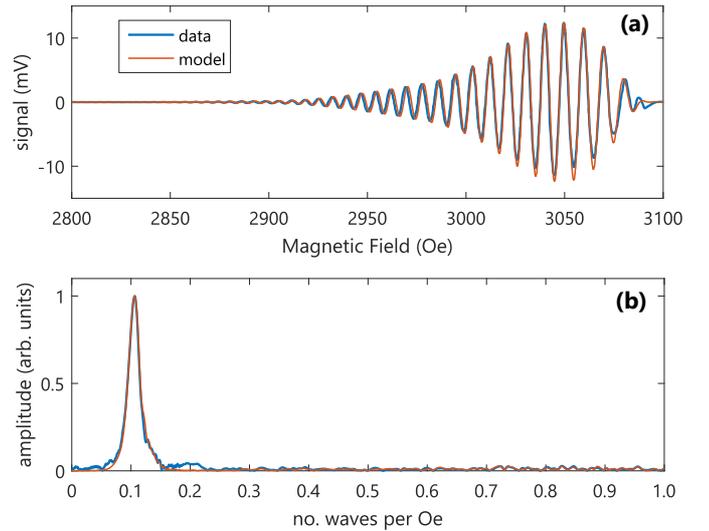}
	\caption{(a) Measured output of lock-in amplifier (blue) compared with calculated output (orange). A full oscillation corresponds to a change of 1 to the number of full wavelengths across the width of the waveguide. (b) The Fourier transform of (a) showing the change in number of waves as a function of magnetic field.  }	
	\label{fig:HomodyneResults}
\end{figure}

The standing wave behavior as a function of applied magnetic field was studied by using a standard homodyne technique \cite{Lim2018} with the set-up shown in Fig. \ref{fig:blockdiagramschematicHomodyneV1}. A $3.915\ghz$ signal was split into two equal parts, one of which was a reference that was fed directly into one port of a mixer. The other signal was used to drive a pump antenna and excite magnons that propagated across the width of the waveguide which were then detected by the second pump antenna. This signal, having picked up a phase difference,  $\Delta\phi=kw$, was then fed into the other mixer port and multiplied with the reference signal. The fast oscillating component of the output was removed with a low-pass filter. The use of the lock-in technique meant that the normally d.c. component was oscillating at the field modulation frequency since the modulation coils were driven by the reference from the SR850 lock-in amplifier. The output signal, depicted in Fig. \ref{fig:HomodyneResults} (a), underwent a complete oscillation for each change in wavenumber $\Delta k = 2\pi/w$, that is, when the number of full wavelengths across the width changes by one.

Figure \ref{fig:HomodyneResults}(a) shows how the lock-in signal oscillates with applied magnetic field. Also shown in orange is a theoretical curve obtained by multiplying $w$ by the wavenumber $k(\omega,H)$ that was calculated for the $n=1$ thickness FVMSW mode. The cosine of this product was then taken and multiplied by the transmission envelope of the data. The data and the model show a good match with oscillations occurring approximately every $10 \Oe$. Fourier transforms of the curves in (a) also show a good correlation and peak at 0.106 waves per $\Oe$. These data demonstrate that as the applied magnetic field is increased by approximately $10\Oe$, the integer number of waves across the waveguide decreases by 1, until the FMR mode is reached. Since a standing wave occurs for a half-integer number of wavelengths, this therefore suggests that a DAC will be supported at intervals of approximately $5\Oe$.

\subsection{Measuring fractal behavior}

The experiment to observe time-domain fractal behavior utilized magnons of two distinct frequencies, excited in separate regions of the waveguide. The probe signal is excited at $f'$ and propagates along the waveguide to the DAC comprising counter-propagating magnons of frequency $f_{\mathrm{DAC}}$.
When the incoming magnons with $f'$ enters the region of periodic potential, the locally-altered dispersion facilitates highly nonlinear effects \cite{Inglis2019} including the formation of a time-domain fractal. 

The external magnetic field was applied normal to the film: a field configuration that is necessary to excite forward volume magnetostatic spin waves. \cite{Prabhakar2009}. The sketch in Fig. \ref{fig:yigsetup} shows the two inputs of the experiment: the parallel antennae acting as pumps to create a standing spin wave of frequency $f_{\mathrm{DAC}}$, and the probe antenna exciting a travelling FVMSW with $f'$. 

\begin{figure}[htbp]
	\centering
		\includegraphics[width=0.4\textwidth]{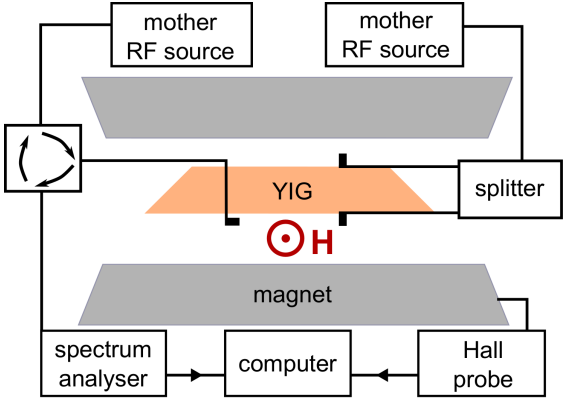}
	\caption{Illustration of experimental apparatus for measuring fractal behavior.}
	
	\label{fig:blockdiagram}
\end{figure}

Figure \ref{fig:blockdiagram} shows a diagram of the experimental equipment used. The pump antennae were driven by a microwave source Hewlett Packard HP8672A with a frequency equal to $f_{\mathrm{DAC}} = 3.915 \ghz$ and power approximately $P_{\mathrm{DAC}}= 13 \mathrm{\, dBm}$. The probe antenna was connected to a circulator allowing for excitation and detection of spin waves with a single antenna. This antenna was driven by a separate microwave source (HP8671A) at $f'=3.917 \ghz$, with an input power of $P' =  0 \mathrm{\, dBm}$. Spin wave detection was performed by a spectrum analyzer. The magnetic field was applied to the sample using an electromagnet that was swept over the region of interest from $3070\Oe$ to $3105\Oe$.

\section{Results and Discussion}

\begin{figure}[htbp]
	\centering
		\includegraphics[width=0.48\textwidth]{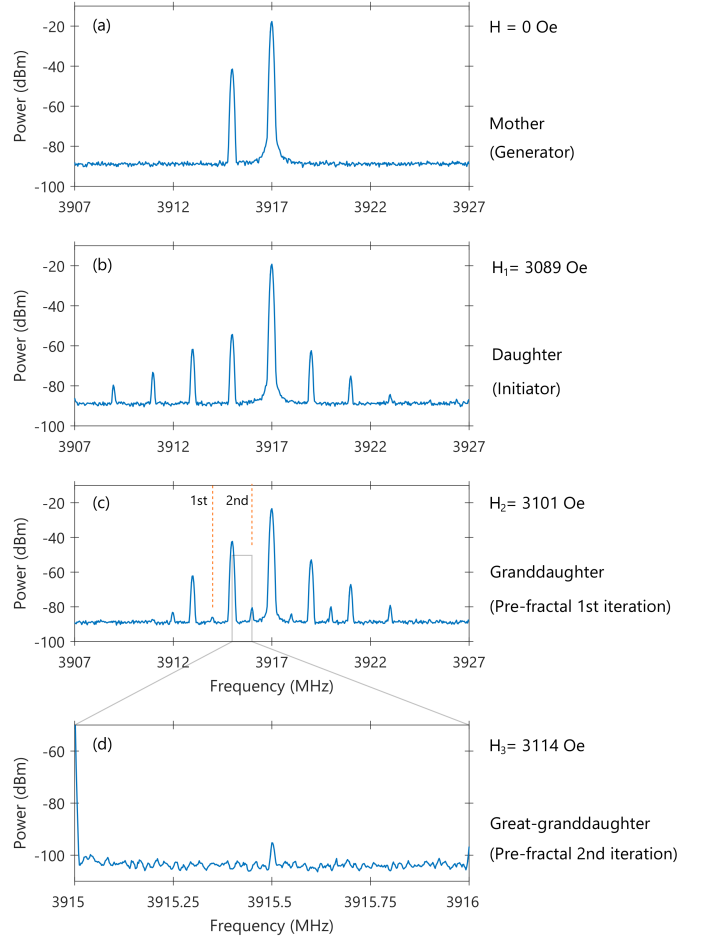}
	\caption{Measured frequency spectra for various magnetic fields.(a) Input frequencies constituting the mother mode. (b) Frequency comb   representing the daughter modes. (c) Generation of granddaughter peaks which appear as intermediary peaks midway between daughters. Orange lines highlight the 1st and 2nd granddaughter fractal modes. (d) A zoomed in spectra showing a great-granddaughter mode between a daughter and granddaughter peak. }	
	\label{fig:FractalTraces}
\end{figure}

Figure \ref{fig:FractalTraces} shows the different frequency spectra for various magnetic fields. The initiator spectrum measured at zero field is shown in Fig. \ref{fig:FractalTraces}(a), comprising the signals $f_{\mathrm{DAC}}$ and $f'$. The measured power $P_{\mathrm{DAC}}$ was the direct coupling between the pump and the signal antenna, while the large $P'$ signal was due to reflections at the probe antenna resulting from impedance mismatch. The magnetic field was increased to $H = 3089 \mathrm{\, Oe}$ resulting in the creation of daughter peaks depicted in \ref{fig:FractalTraces}(b). This comb-like pattern appears with intervals of $\Delta f^{(0)} = \mid f_{\mathrm{DAC}} - f'\mid = 2 \mhz$, forming the structure that is repeated on successively smaller scales. Also note that the amplitude of the peak at $3.915\ghz$ has decreased upon generation of daughter modes and has been distributed to the surrounding comb. As the field was increased again to $H = 3101 \mathrm{\, Oe}$ new granddaughter modes appeared between the mother and daughter peaks shown in (c), where the orange dashed lines highlight the 1st and 2nd granddaughter modes that straddle the $f_{\mathrm{DAC}}$ signal. At this magnetic field strength, the frequency interval between peaks has been reduced to $\Delta f^{(0)}/2 = \Delta f^{(1)} = 1 \mhz$. The magnetic field was increased further to $H = 3114 \mathrm{\, Oe}$ the results of which are shown in Fig. \ref{fig:FractalTraces}(d). A great-granddaughter peak was observed with at 3.9155 GHz, between the mother and the 2nd granddaughter peak, boxed in \ref{fig:FractalTraces}(c). Significantly, the great-granddaughter peak appears at a frequency interval of $\Delta f^{(0)}/4 = \Delta f^{(1)}/2 = \Delta f^{(2)} = 0.5 \mhz$. These data suggest the onset of a fractal time-domain structure. While power limitations were set by the operating threshold of the equipment, we posit that higher powers would result in the observation of a 3rd iteration of pre-fractal, following a pattern wherein the $n$th pre-fractal observed would be a frequency comb of interval $\Delta f^{(n)} = \Delta f^{(0)}/2^{n} $.

The observed peaks were extremely sensitive in response to magnetic field strength. The appearance and disappearance of the peaks was often abrupt, with 1st and 2nd iteration pre-fractals occurring only over a tight field range. 

The results show a period-doubling bifurcation similar to that observed by a high power CW monochromatic input to undergo self modulational instability (SMI) resulting in a comb of frequnecies, corresponding to a soliton pulse train. As the input power was increased, the frequency interval of the comb was spontaneously halved, doubling the period of the soliton train \cite{Wu2004}. While previous reports have only observed one round of period-doubling, our observation of a successive stage of period-doubling, coupled with presence of the periodic potential of the DAC, suggest the onset of time-domain fractals. 

The mechanism responsible for the generation of daughter peaks can be explained by four-wave mixing (FWM) \cite{Khivintsev2011,Inglis2019}. For the physical process to occur, the following energy and momentum conservation condition must be satisfied: 
\begin{equation}
f_{1} + f_{2} = f_{3} + f_{4}.
\label{consenergy}
\end{equation}
\begin{equation}
\mathbf{k}_{1} + \mathbf{k}_{1} = \mathbf{k}_{3} + \mathbf{k}_{4}.
\label{consmom}
\end{equation}
In the case of our experiment we may rewrite eqn.(\ref{consenergy}) as $2f_{\mathrm{DAC}} = f' + (f_{\mathrm{DAC}}+\Delta f^{(0)})$, where the final term in parentheses is the first daughter mode to the right of $f_{\mathrm{DAC}}$. Similarly, we may obtain the first daughter mode to the left of $f'$ by rewriting eqn.(\ref{consenergy}) as $2f' = f_{\mathrm{DAC}} + (f'-\Delta f^{(0)})$. Once we have the first set of sidebands the method may be repeated to obtain the entire comb of daughter peaks. This type of four magnon scattering is a physical process that is to be expected and is well understood and documented in the literature. While we also suggest that FWM is responsible for the generation of granddaughter and great-granddaughter peaks, the specific origin of the well-defined pattern is not completely understood. 
\begin{figure}[htbp]
	\centering
		\includegraphics[width=0.48\textwidth]{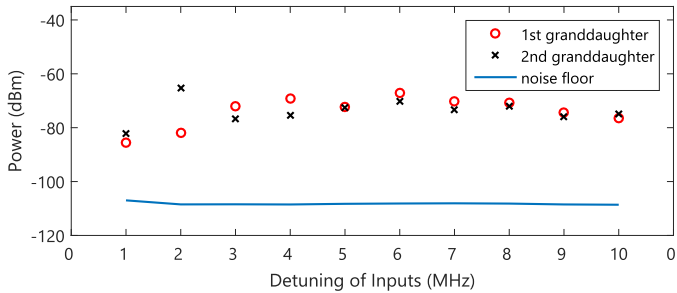}
	\caption{Amplitude of granddaughter modes as a function of detuning $\Delta f^{(0)}$. Noise floor for each measurement was approximately $-110\dbm$. }	
	\label{fig:FractalDetuning}
\end{figure}

\subsection{Detuning effects}
Previous observations of period-doubling and fractals \cite{Wu2004,Richardson2018} have occurred as a result of SMI. Our experiment differs from these reports in that the mother mode consists of two distinct frequencies and therefore the daughter modes were generated using induced modulational instability (IMI) \cite{Demidov1998}. 

The effect the mother mode detuning on the fractal amplitude was  investigated.  Figure \ref{fig:FractalDetuning} shows how the maximum amplitude of the 1st and 2nd granddaughter peaks vary as the detuning $\Delta f^{(0)}$ changes from  $1-10\mhz$, in steps of $1\mhz$. In this experiment $f_{\mathrm{DAC}}$ was fixed at $3.915\ghz$, while $f'$ was varied between $3.916\ghz$ and $3.925\ghz$. For each $\Delta f^{(0)}$ two field sweeps were performed, with the spectrum analyser centred on the 1st and 2nd granddaughter, respectively. The span was decreased to $80\khz$ to increase the sensitivity and drop the noise floor to approximately $-110\dbm$. These data show that for each $\Delta f^{(0)}$ the two granddaughter peaks have similar power levels. Furthermore, the amplitude of the peaks is fairly robust as the detuning is above $2\mhz$. 

\begin{figure*}[htbp]
	\centering
		\includegraphics[width=0.98\textwidth]{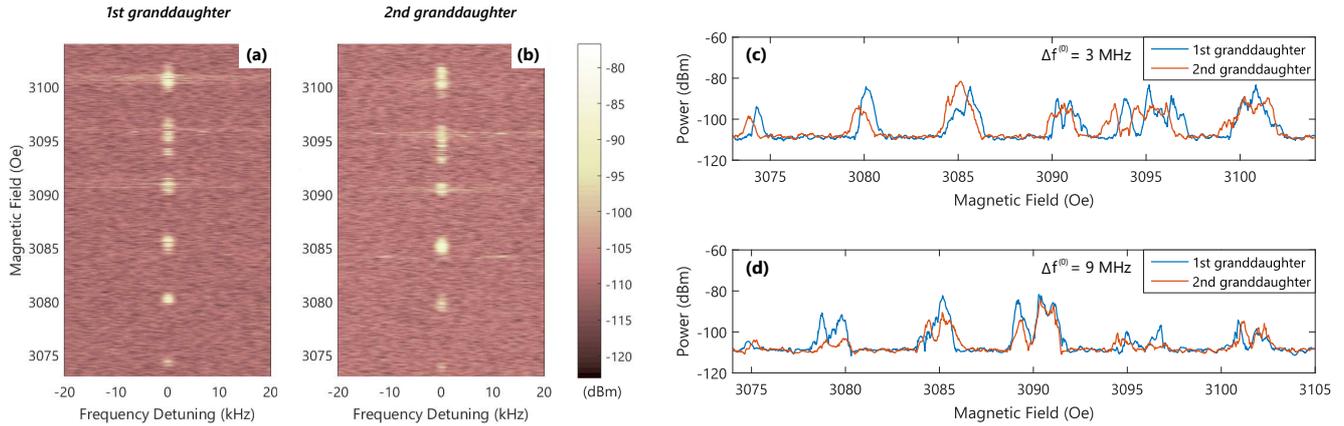}
	\caption{(a),(b) Field dependence of spectra centred on the 1st and 2nd granddaughter mode, respectively, for $\Delta f^{(0)}=3\mhz$. (c),(d) field dependence of granddaughter modes for $\Delta f^{(0)}=3\mhz$ and $9\mhz$.}	
	\label{fig:FractalsDetuningAnalysis}
\end{figure*}

The field dependence of the 1st and 2nd granddaughter peaks for a specific detuning $\Delta f^{(0)}=3\mhz$ is also presented in Figure \ref{fig:FractalsDetuningAnalysis}. Subfigures (a) and (b) show a series of spectra centred on the predicted frequencies of the 1st and 2nd granddaughters: $3913.5\mhz$ and $3916.5\mhz$, respectively. It is clear that these fractal modes only appear for specific magnetic fields. This dependence has a clear periodicity which may be seen in  (c), where the grandduaghter signals have been plotted with respect to field only. A similar trace is presented in (d) for a detuning of $\Delta f^{(0)}=3\mhz$. We see that there is a signal occurring approximately every $5\Oe$, which corresponds to the field dependence of the DAC, suggesting that the signal is a result of the spatio-temporally periodic texture of the dynamic artificial crystal. The peaks become less well defined and have greater spectral width as the magnetic field increases, resembling the onset of chaotic effects \cite{Wu2009}. 

\section{Conclusion}
In conclusion, we have observed experimentally for the first time, spontaneous fractals in simple un-patterned thin film. Specifically, we utilized an all-magnon dynamic artificial crystal (DAC) to observe  nonlinear spin wave dynamics, resulting in the onset of a time-domain exact fractal pattern. We demonstrate successive levels of modulation with the each new set of side-bands appearing half-way between the previous. The extension of this pattern is that the $n$th pre-fractal has an interval of $\Delta f^{(0)}/2^{n}$. Furthermore, we have observed that this pattern is robust over a range of frequency detunings. Our observations differ from previous studies in that the film is not lithographed and the fractal pattern was generated using IMI allowing a tunable period-doubling effect. Not only does this offering greater control for use in any potential spin wave devices, but also provides a further insight into fundamental spin wave dynamics.

We would like to thank Prof. Paul Ewart for his invaluable insight in developing this experiment. This research was partially funded by Magdalen College, Oxford.

\bibliographystyle{elsarticle-num}
\bibliography{FractalPaper}

\begin{thebibliography}{32}%
\makeatletter
\providecommand \@ifxundefined [1]{%
 \@ifx{#1\undefined}
}%
\providecommand \@ifnum [1]{%
 \ifnum #1\expandafter \@firstoftwo
 \else \expandafter \@secondoftwo
 \fi
}%
\providecommand \@ifx [1]{%
 \ifx #1\expandafter \@firstoftwo
 \else \expandafter \@secondoftwo
 \fi
}%
\providecommand \natexlab [1]{#1}%
\providecommand \enquote  [1]{``#1''}%
\providecommand \bibnamefont  [1]{#1}%
\providecommand \bibfnamefont [1]{#1}%
\providecommand \citenamefont [1]{#1}%
\providecommand \href@noop [0]{\@secondoftwo}%
\providecommand \href [0]{\begingroup \@sanitize@url \@href}%
\providecommand \@href[1]{\@@startlink{#1}\@@href}%
\providecommand \@@href[1]{\endgroup#1\@@endlink}%
\providecommand \@sanitize@url [0]{\catcode `\\12\catcode `\$12\catcode
  `\&12\catcode `\#12\catcode `\^12\catcode `\_12\catcode `\%12\relax}%
\providecommand \@@startlink[1]{}%
\providecommand \@@endlink[0]{}%
\providecommand \url  [0]{\begingroup\@sanitize@url \@url }%
\providecommand \@url [1]{\endgroup\@href {#1}{\urlprefix }}%
\providecommand \urlprefix  [0]{URL }%
\providecommand \Eprint [0]{\href }%
\providecommand \doibase [0]{http://dx.doi.org/}%
\providecommand \selectlanguage [0]{\@gobble}%
\providecommand \bibinfo  [0]{\@secondoftwo}%
\providecommand \bibfield  [0]{\@secondoftwo}%
\providecommand \translation [1]{[#1]}%
\providecommand \BibitemOpen [0]{}%
\providecommand \bibitemStop [0]{}%
\providecommand \bibitemNoStop [0]{.\EOS\space}%
\providecommand \EOS [0]{\spacefactor3000\relax}%
\providecommand \BibitemShut  [1]{\csname bibitem#1\endcsname}%
\let\auto@bib@innerbib\@empty
\bibitem [{\citenamefont {Addison}(1997)}]{Addison1997}%
  \BibitemOpen
  \bibfield  {author} {\bibinfo {author} {\bibfnamefont {P.}~\bibnamefont
  {Addison}},\ }\href@noop {} {\emph {\bibinfo {title} {{Fractals and Chaos: An
  Illustrated Course}}}}\ (\bibinfo  {publisher} {Bristol: Institute of Physics
  Publishing.},\ \bibinfo {year} {1997})\BibitemShut {NoStop}%
\bibitem [{\citenamefont {Mandelbrot}(1984)}]{Mandelbrot1984}%
  \BibitemOpen
  \bibfield  {author} {\bibinfo {author} {\bibfnamefont {B.~B.}\ \bibnamefont
  {Mandelbrot}},\ }\href@noop {} {\emph {\bibinfo {title} {{The Fractal
  Geometry of Nature}}}}\ (\bibinfo  {publisher} {W.H. Freeman},\ \bibinfo
  {address} {New York},\ \bibinfo {year} {1984})\BibitemShut {NoStop}%
\bibitem [{\citenamefont {Taylor}\ \emph {et~al.}(2007)\citenamefont {Taylor},
  \citenamefont {Guzman}, \citenamefont {Martin}, \citenamefont {Hall},
  \citenamefont {Micolich}, \citenamefont {Jonas}, \citenamefont {Scannell},
  \citenamefont {Fairbanks},\ and\ \citenamefont {Marlow}}]{Taylor2007}%
  \BibitemOpen
  \bibfield  {author} {\bibinfo {author} {\bibfnamefont {R.}~\bibnamefont
  {Taylor}}, \bibinfo {author} {\bibfnamefont {R.}~\bibnamefont {Guzman}},
  \bibinfo {author} {\bibfnamefont {T.}~\bibnamefont {Martin}}, \bibinfo
  {author} {\bibfnamefont {G.}~\bibnamefont {Hall}}, \bibinfo {author}
  {\bibfnamefont {A.}~\bibnamefont {Micolich}}, \bibinfo {author}
  {\bibfnamefont {D.}~\bibnamefont {Jonas}}, \bibinfo {author} {\bibfnamefont
  {B.}~\bibnamefont {Scannell}}, \bibinfo {author} {\bibfnamefont
  {M.}~\bibnamefont {Fairbanks}}, \ and\ \bibinfo {author} {\bibfnamefont
  {C.}~\bibnamefont {Marlow}},\ }\href {\doibase 10.1016/j.patrec.2006.08.012}
  {\bibfield  {journal} {\bibinfo  {journal} {Pattern Recognit. Lett.}\
  }\textbf {\bibinfo {volume} {28}},\ \bibinfo {pages} {695} (\bibinfo {year}
  {2007})}\BibitemShut {NoStop}%
\bibitem [{\citenamefont {Segev}\ \emph {et~al.}(2012)\citenamefont {Segev},
  \citenamefont {Solja{\v{c}}i{\'{c}}},\ and\ \citenamefont
  {Dudley}}]{Segev2012}%
  \BibitemOpen
  \bibfield  {author} {\bibinfo {author} {\bibfnamefont {M.}~\bibnamefont
  {Segev}}, \bibinfo {author} {\bibfnamefont {M.}~\bibnamefont
  {Solja{\v{c}}i{\'{c}}}}, \ and\ \bibinfo {author} {\bibfnamefont {J.~M.}\
  \bibnamefont {Dudley}},\ }\href {\doibase 10.1038/nphoton.2012.71} {\bibfield
   {journal} {\bibinfo  {journal} {Nat. Photonics}\ }\textbf {\bibinfo {volume}
  {6}},\ \bibinfo {pages} {209} (\bibinfo {year} {2012})}\BibitemShut {NoStop}%
\bibitem [{\citenamefont {Soljacic}\ \emph {et~al.}(2000)\citenamefont
  {Soljacic}, \citenamefont {Segev},\ and\ \citenamefont
  {Menyuk}}]{Soljacic2000}%
  \BibitemOpen
  \bibfield  {author} {\bibinfo {author} {\bibfnamefont {M.}~\bibnamefont
  {Soljacic}}, \bibinfo {author} {\bibfnamefont {M.}~\bibnamefont {Segev}}, \
  and\ \bibinfo {author} {\bibfnamefont {C.~R.}\ \bibnamefont {Menyuk}},\
  }\href {\doibase 10.1103/PhysRevE.61.R1048} {\bibfield  {journal} {\bibinfo
  {journal} {Phys. Rev. E}\ }\textbf {\bibinfo {volume} {61}},\ \bibinfo
  {pages} {R1048} (\bibinfo {year} {2000})}\BibitemShut {NoStop}%
\bibitem [{\citenamefont {Sears}\ \emph {et~al.}(2000)\citenamefont {Sears},
  \citenamefont {Soljacic}, \citenamefont {Segev}, \citenamefont {Krylov},\
  and\ \citenamefont {Bergman}}]{Sears2000}%
  \BibitemOpen
  \bibfield  {author} {\bibinfo {author} {\bibfnamefont {S.}~\bibnamefont
  {Sears}}, \bibinfo {author} {\bibfnamefont {M.}~\bibnamefont {Soljacic}},
  \bibinfo {author} {\bibfnamefont {M.}~\bibnamefont {Segev}}, \bibinfo
  {author} {\bibfnamefont {D.}~\bibnamefont {Krylov}}, \ and\ \bibinfo {author}
  {\bibfnamefont {K.}~\bibnamefont {Bergman}},\ }\href {\doibase
  10.1103/PhysRevLett.84.1902} {\bibfield  {journal} {\bibinfo  {journal}
  {Phys. Rev. Lett.}\ }\textbf {\bibinfo {volume} {84}},\ \bibinfo {pages}
  {1902} (\bibinfo {year} {2000})}\BibitemShut {NoStop}%
\bibitem [{\citenamefont {Erkintalo}\ \emph {et~al.}(2012)\citenamefont
  {Erkintalo}, \citenamefont {Hammani}, \citenamefont {Kibler}, \citenamefont
  {Finot}, \citenamefont {Akhmediev}, \citenamefont {Dudley},\ and\
  \citenamefont {Genty}}]{Erkintalo2012}%
  \BibitemOpen
  \bibfield  {author} {\bibinfo {author} {\bibfnamefont {M.}~\bibnamefont
  {Erkintalo}}, \bibinfo {author} {\bibfnamefont {K.}~\bibnamefont {Hammani}},
  \bibinfo {author} {\bibfnamefont {B.}~\bibnamefont {Kibler}}, \bibinfo
  {author} {\bibfnamefont {C.}~\bibnamefont {Finot}}, \bibinfo {author}
  {\bibfnamefont {N.}~\bibnamefont {Akhmediev}}, \bibinfo {author}
  {\bibfnamefont {J.~M.}\ \bibnamefont {Dudley}}, \ and\ \bibinfo {author}
  {\bibfnamefont {G.}~\bibnamefont {Genty}},\ }\href {\doibase
  10.1109/ICTON.2012.6253900} {\bibfield  {journal} {\bibinfo  {journal} {Int.
  Conf. Transparent Opt. Networks}\ }\textbf {\bibinfo {volume} {253901}},\
  \bibinfo {pages} {14} (\bibinfo {year} {2012})}\BibitemShut {NoStop}%
\bibitem [{\citenamefont {Wu}\ \emph {et~al.}(2006)\citenamefont {Wu},
  \citenamefont {Kalinikos}, \citenamefont {Carr},\ and\ \citenamefont
  {Patton}}]{Wu2006}%
  \BibitemOpen
  \bibfield  {author} {\bibinfo {author} {\bibfnamefont {M.}~\bibnamefont
  {Wu}}, \bibinfo {author} {\bibfnamefont {B.~A.}\ \bibnamefont {Kalinikos}},
  \bibinfo {author} {\bibfnamefont {L.~D.}\ \bibnamefont {Carr}}, \ and\
  \bibinfo {author} {\bibfnamefont {C.~E.}\ \bibnamefont {Patton}},\ }\href
  {\doibase 10.1103/PhysRevLett.96.187202} {\bibfield  {journal} {\bibinfo
  {journal} {Phys. Rev. Lett.}\ }\textbf {\bibinfo {volume} {96}},\ \bibinfo
  {pages} {187202} (\bibinfo {year} {2006})}\BibitemShut {NoStop}%
\bibitem [{\citenamefont {Richardson}\ \emph {et~al.}(2018)\citenamefont
  {Richardson}, \citenamefont {Kalinikos}, \citenamefont {Carr},\ and\
  \citenamefont {Wu}}]{Richardson2018}%
  \BibitemOpen
  \bibfield  {author} {\bibinfo {author} {\bibfnamefont {D.}~\bibnamefont
  {Richardson}}, \bibinfo {author} {\bibfnamefont {B.~A.}\ \bibnamefont
  {Kalinikos}}, \bibinfo {author} {\bibfnamefont {L.~D.}\ \bibnamefont {Carr}},
  \ and\ \bibinfo {author} {\bibfnamefont {M.}~\bibnamefont {Wu}},\ }\href
  {\doibase 10.1103/PhysRevLett.121.107204} {\bibfield  {journal} {\bibinfo
  {journal} {Phys. Rev. Lett.}\ }\textbf {\bibinfo {volume} {121}},\ \bibinfo
  {pages} {107204} (\bibinfo {year} {2018})}\BibitemShut {NoStop}%
\bibitem [{\citenamefont {Ord{\'{o}}{\~{n}}ez-Romero}\ \emph
  {et~al.}(2016)\citenamefont {Ord{\'{o}}{\~{n}}ez-Romero}, \citenamefont
  {Lazcano-Ortiz}, \citenamefont {Drozdovskii}, \citenamefont {Kalinikos},
  \citenamefont {Aguilar-Huerta}, \citenamefont {Dom{\'{i}}nguez-Ju{\'{a}}rez},
  \citenamefont {Lopez-Maldonado}, \citenamefont {Qureshi}, \citenamefont
  {Kolokoltsev},\ and\ \citenamefont {Monsivais}}]{Ordonez-Romero2016}%
  \BibitemOpen
  \bibfield  {author} {\bibinfo {author} {\bibfnamefont {C.~L.}\ \bibnamefont
  {Ord{\'{o}}{\~{n}}ez-Romero}}, \bibinfo {author} {\bibfnamefont
  {Z.}~\bibnamefont {Lazcano-Ortiz}}, \bibinfo {author} {\bibfnamefont
  {A.}~\bibnamefont {Drozdovskii}}, \bibinfo {author} {\bibfnamefont
  {B.}~\bibnamefont {Kalinikos}}, \bibinfo {author} {\bibfnamefont
  {M.}~\bibnamefont {Aguilar-Huerta}}, \bibinfo {author} {\bibfnamefont
  {J.~L.}\ \bibnamefont {Dom{\'{i}}nguez-Ju{\'{a}}rez}}, \bibinfo {author}
  {\bibfnamefont {G.}~\bibnamefont {Lopez-Maldonado}}, \bibinfo {author}
  {\bibfnamefont {N.}~\bibnamefont {Qureshi}}, \bibinfo {author} {\bibfnamefont
  {O.}~\bibnamefont {Kolokoltsev}}, \ and\ \bibinfo {author} {\bibfnamefont
  {G.}~\bibnamefont {Monsivais}},\ }\href {\doibase 10.1063/1.4958903}
  {\bibfield  {journal} {\bibinfo  {journal} {J. Appl. Phys.}\ }\textbf
  {\bibinfo {volume} {120}},\ \bibinfo {pages} {043901} (\bibinfo {year}
  {2016})}\BibitemShut {NoStop}%
\bibitem [{\citenamefont {Zakharov}\ and\ \citenamefont
  {Ostrovsky}(2009)}]{Zakharov2009}%
  \BibitemOpen
  \bibfield  {author} {\bibinfo {author} {\bibfnamefont {V.~E.}\ \bibnamefont
  {Zakharov}}\ and\ \bibinfo {author} {\bibfnamefont {L.~A.}\ \bibnamefont
  {Ostrovsky}},\ }\href {\doibase 10.1016/j.physd.2008.12.002} {\bibfield
  {journal} {\bibinfo  {journal} {Phys. D Nonlinear Phenom.}\ }\textbf
  {\bibinfo {volume} {238}},\ \bibinfo {pages} {540} (\bibinfo {year}
  {2009})}\BibitemShut {NoStop}%
\bibitem [{\citenamefont {Karenowska}\ \emph {et~al.}(2012)\citenamefont
  {Karenowska}, \citenamefont {Gregg}, \citenamefont {Tiberkevich},
  \citenamefont {Slavin}, \citenamefont {Chumak}, \citenamefont {Serga},\ and\
  \citenamefont {Hillebrands}}]{Karenowska2012}%
  \BibitemOpen
  \bibfield  {author} {\bibinfo {author} {\bibfnamefont {A.~D.}\ \bibnamefont
  {Karenowska}}, \bibinfo {author} {\bibfnamefont {J.~F.}\ \bibnamefont
  {Gregg}}, \bibinfo {author} {\bibfnamefont {V.~S.}\ \bibnamefont
  {Tiberkevich}}, \bibinfo {author} {\bibfnamefont {A.~N.}\ \bibnamefont
  {Slavin}}, \bibinfo {author} {\bibfnamefont {A.~V.}\ \bibnamefont {Chumak}},
  \bibinfo {author} {\bibfnamefont {A.~A.}\ \bibnamefont {Serga}}, \ and\
  \bibinfo {author} {\bibfnamefont {B.}~\bibnamefont {Hillebrands}},\ }\href
  {\doibase 10.1103/PhysRevLett.108.015505} {\bibfield  {journal} {\bibinfo
  {journal} {Phys. Rev. Lett.}\ }\textbf {\bibinfo {volume} {108}},\ \bibinfo
  {pages} {015505} (\bibinfo {year} {2012})},\ \Eprint
  {http://arxiv.org/abs/1106.3722} {arXiv:1106.3722} \BibitemShut {NoStop}%
\bibitem [{\citenamefont {Chumak}\ \emph {et~al.}(2010)\citenamefont {Chumak},
  \citenamefont {Tiberkevich}, \citenamefont {Karenowska}, \citenamefont
  {Serga}, \citenamefont {Gregg}, \citenamefont {Slavin},\ and\ \citenamefont
  {Hillebrands}}]{Chumak2010a}%
  \BibitemOpen
  \bibfield  {author} {\bibinfo {author} {\bibfnamefont {A.~V.}\ \bibnamefont
  {Chumak}}, \bibinfo {author} {\bibfnamefont {V.~S.}\ \bibnamefont
  {Tiberkevich}}, \bibinfo {author} {\bibfnamefont {A.~D.}\ \bibnamefont
  {Karenowska}}, \bibinfo {author} {\bibfnamefont {A.~a.}\ \bibnamefont
  {Serga}}, \bibinfo {author} {\bibfnamefont {J.~F.}\ \bibnamefont {Gregg}},
  \bibinfo {author} {\bibfnamefont {A.~N.}\ \bibnamefont {Slavin}}, \ and\
  \bibinfo {author} {\bibfnamefont {B.}~\bibnamefont {Hillebrands}},\ }\href
  {\doibase 10.1038/ncomms1142} {\bibfield  {journal} {\bibinfo  {journal}
  {Nat. Commun.}\ }\textbf {\bibinfo {volume} {1}},\ \bibinfo {pages} {141}
  (\bibinfo {year} {2010})}\BibitemShut {NoStop}%
\bibitem [{\citenamefont {Schneider}\ \emph {et~al.}(2010)\citenamefont
  {Schneider}, \citenamefont {Serga}, \citenamefont {Chumak}, \citenamefont
  {Sandweg}, \citenamefont {Trudel}, \citenamefont {Wolff}, \citenamefont
  {Kostylev}, \citenamefont {Tiberkevich}, \citenamefont {Slavin},\ and\
  \citenamefont {Hillebrands}}]{Schneider2010}%
  \BibitemOpen
  \bibfield  {author} {\bibinfo {author} {\bibfnamefont {T.}~\bibnamefont
  {Schneider}}, \bibinfo {author} {\bibfnamefont {A.~A.}\ \bibnamefont
  {Serga}}, \bibinfo {author} {\bibfnamefont {A.~V.}\ \bibnamefont {Chumak}},
  \bibinfo {author} {\bibfnamefont {C.~W.}\ \bibnamefont {Sandweg}}, \bibinfo
  {author} {\bibfnamefont {S.}~\bibnamefont {Trudel}}, \bibinfo {author}
  {\bibfnamefont {S.}~\bibnamefont {Wolff}}, \bibinfo {author} {\bibfnamefont
  {M.~P.}\ \bibnamefont {Kostylev}}, \bibinfo {author} {\bibfnamefont {V.~S.}\
  \bibnamefont {Tiberkevich}}, \bibinfo {author} {\bibfnamefont {A.~N.}\
  \bibnamefont {Slavin}}, \ and\ \bibinfo {author} {\bibfnamefont
  {B.}~\bibnamefont {Hillebrands}},\ }\href {\doibase
  10.1103/PhysRevLett.104.197203} {\bibfield  {journal} {\bibinfo  {journal}
  {Phys. Rev. Lett.}\ }\textbf {\bibinfo {volume} {104}},\ \bibinfo {pages}
  {197203} (\bibinfo {year} {2010})}\BibitemShut {NoStop}%
\bibitem [{\citenamefont {Demidov}\ \emph
  {et~al.}(2009{\natexlab{a}})\citenamefont {Demidov}, \citenamefont
  {Demokritov}, \citenamefont {Birt}, \citenamefont {O'Gorman}, \citenamefont
  {Tsoi},\ and\ \citenamefont {Li}}]{Demidov2009c}%
  \BibitemOpen
  \bibfield  {author} {\bibinfo {author} {\bibfnamefont {V.~E.}\ \bibnamefont
  {Demidov}}, \bibinfo {author} {\bibfnamefont {S.~O.}\ \bibnamefont
  {Demokritov}}, \bibinfo {author} {\bibfnamefont {D.}~\bibnamefont {Birt}},
  \bibinfo {author} {\bibfnamefont {B.}~\bibnamefont {O'Gorman}}, \bibinfo
  {author} {\bibfnamefont {M.}~\bibnamefont {Tsoi}}, \ and\ \bibinfo {author}
  {\bibfnamefont {X.}~\bibnamefont {Li}},\ }\href {\doibase
  10.1103/PhysRevB.80.014429} {\bibfield  {journal} {\bibinfo  {journal} {Phys.
  Rev. B}\ }\textbf {\bibinfo {volume} {80}},\ \bibinfo {pages} {014429}
  (\bibinfo {year} {2009}{\natexlab{a}})}\BibitemShut {NoStop}%
\bibitem [{\citenamefont {Kalinikos}\ and\ \citenamefont
  {Slavin}(1986)}]{Kalinikos1986a}%
  \BibitemOpen
  \bibfield  {author} {\bibinfo {author} {\bibfnamefont {B.~A.}\ \bibnamefont
  {Kalinikos}}\ and\ \bibinfo {author} {\bibfnamefont {A.~N.}\ \bibnamefont
  {Slavin}},\ }\href {\doibase 10.1088/0022-3719/19/35/014} {\bibfield
  {journal} {\bibinfo  {journal} {J. Phys. C Solid State Phys.}\ }\textbf
  {\bibinfo {volume} {19}},\ \bibinfo {pages} {7013} (\bibinfo {year}
  {1986})}\BibitemShut {NoStop}%
\bibitem [{\citenamefont {Boardman}\ \emph {et~al.}(1995)\citenamefont
  {Boardman}, \citenamefont {Nikitov}, \citenamefont {Xie},\ and\ \citenamefont
  {Mehta}}]{Boardman1995}%
  \BibitemOpen
  \bibfield  {author} {\bibinfo {author} {\bibfnamefont {A.}~\bibnamefont
  {Boardman}}, \bibinfo {author} {\bibfnamefont {S.}~\bibnamefont {Nikitov}},
  \bibinfo {author} {\bibfnamefont {K.}~\bibnamefont {Xie}}, \ and\ \bibinfo
  {author} {\bibfnamefont {H.}~\bibnamefont {Mehta}},\ }\href {\doibase
  10.1016/0304-8853(94)01624-0} {\bibfield  {journal} {\bibinfo  {journal} {J.
  Magn. Magn. Mater.}\ }\textbf {\bibinfo {volume} {145}},\ \bibinfo {pages}
  {357} (\bibinfo {year} {1995})}\BibitemShut {NoStop}%
\bibitem [{\citenamefont {Demidov}\ \emph
  {et~al.}(2009{\natexlab{b}})\citenamefont {Demidov}, \citenamefont
  {Kostylev}, \citenamefont {Rott}, \citenamefont {Krzysteczko}, \citenamefont
  {Reiss},\ and\ \citenamefont {Demokritov}}]{Demidov2009b}%
  \BibitemOpen
  \bibfield  {author} {\bibinfo {author} {\bibfnamefont {V.~E.}\ \bibnamefont
  {Demidov}}, \bibinfo {author} {\bibfnamefont {M.~P.}\ \bibnamefont
  {Kostylev}}, \bibinfo {author} {\bibfnamefont {K.}~\bibnamefont {Rott}},
  \bibinfo {author} {\bibfnamefont {P.}~\bibnamefont {Krzysteczko}}, \bibinfo
  {author} {\bibfnamefont {G.}~\bibnamefont {Reiss}}, \ and\ \bibinfo {author}
  {\bibfnamefont {S.~O.}\ \bibnamefont {Demokritov}},\ }\href {\doibase
  10.1063/1.3231875} {\bibfield  {journal} {\bibinfo  {journal} {Appl. Phys.
  Lett.}\ }\textbf {\bibinfo {volume} {95}},\ \bibinfo {pages} {10} (\bibinfo
  {year} {2009}{\natexlab{b}})}\BibitemShut {NoStop}%
\bibitem [{\citenamefont {Phillips}\ and\ \citenamefont
  {Rosenberg}(1966)}]{Phillips1966}%
  \BibitemOpen
  \bibfield  {author} {\bibinfo {author} {\bibfnamefont {T.~G.}\ \bibnamefont
  {Phillips}}\ and\ \bibinfo {author} {\bibfnamefont {H.~M.}\ \bibnamefont
  {Rosenberg}},\ }\href {\doibase 10.1088/0034-4885/29/1/307} {\bibfield
  {journal} {\bibinfo  {journal} {Reports Prog. Phys.}\ }\textbf {\bibinfo
  {volume} {29}},\ \bibinfo {pages} {307} (\bibinfo {year} {1966})}\BibitemShut
  {NoStop}%
\bibitem [{\citenamefont {Bloom}\ and\ \citenamefont
  {Bjorklund}(1977)}]{Bloom1977a}%
  \BibitemOpen
  \bibfield  {author} {\bibinfo {author} {\bibfnamefont {D.~M.}\ \bibnamefont
  {Bloom}}\ and\ \bibinfo {author} {\bibfnamefont {G.~C.}\ \bibnamefont
  {Bjorklund}},\ }\href {\doibase 10.1063/1.89791} {\bibfield  {journal}
  {\bibinfo  {journal} {Appl. Phys. Lett.}\ }\textbf {\bibinfo {volume} {31}},\
  \bibinfo {pages} {592} (\bibinfo {year} {1977})}\BibitemShut {NoStop}%
\bibitem [{\citenamefont {Temiryazev}(1989)}]{Temiryazev1989}%
  \BibitemOpen
  \bibfield  {author} {\bibinfo {author} {\bibfnamefont {A.}~\bibnamefont
  {Temiryazev}},\ }\href@noop {} {\bibfield  {journal} {\bibinfo  {journal}
  {Pis'ma Zh. Eksp. Teor. Fiz.}\ }\textbf {\bibinfo {volume} {50}},\ \bibinfo
  {pages} {202} (\bibinfo {year} {1989})}\BibitemShut {NoStop}%
\bibitem [{\citenamefont {Nikitov}\ \emph {et~al.}(1995)\citenamefont
  {Nikitov}, \citenamefont {Jun}, \citenamefont {Marcelli},\ and\ \citenamefont
  {{De Gasperis}}}]{Nikitov1995}%
  \BibitemOpen
  \bibfield  {author} {\bibinfo {author} {\bibfnamefont {S.~A.}\ \bibnamefont
  {Nikitov}}, \bibinfo {author} {\bibfnamefont {S.}~\bibnamefont {Jun}},
  \bibinfo {author} {\bibfnamefont {R.}~\bibnamefont {Marcelli}}, \ and\
  \bibinfo {author} {\bibfnamefont {P.}~\bibnamefont {{De Gasperis}}},\ }\href
  {\doibase 10.1016/0304-8853(94)01699-2} {\bibfield  {journal} {\bibinfo
  {journal} {J. Magn. Magn. Mater.}\ }\textbf {\bibinfo {volume} {145}}
  (\bibinfo {year} {1995}),\ 10.1016/0304-8853(94)01699-2}\BibitemShut
  {NoStop}%
\bibitem [{\citenamefont {Slavin}\ \emph {et~al.}(2002)\citenamefont {Slavin},
  \citenamefont {Demokritov},\ and\ \citenamefont {Hillebrands}}]{Slavin2002}%
  \BibitemOpen
  \bibfield  {author} {\bibinfo {author} {\bibfnamefont {A.~N.}\ \bibnamefont
  {Slavin}}, \bibinfo {author} {\bibfnamefont {S.~O.}\ \bibnamefont
  {Demokritov}}, \ and\ \bibinfo {author} {\bibfnamefont {B.}~\bibnamefont
  {Hillebrands}},\ }in\ \href@noop {} {\emph {\bibinfo {booktitle} {Spin Dyn.
  Confin. Magn. Struct. I}}}\ (\bibinfo {year} {2002})\BibitemShut {NoStop}%
\bibitem [{\citenamefont {Drozdovskii}\ \emph {et~al.}(2016)\citenamefont
  {Drozdovskii}, \citenamefont {Kalinikos}, \citenamefont {Ustinov},\ and\
  \citenamefont {Stashkevich}}]{Drozdovskii2016}%
  \BibitemOpen
  \bibfield  {author} {\bibinfo {author} {\bibfnamefont {A.~V.}\ \bibnamefont
  {Drozdovskii}}, \bibinfo {author} {\bibfnamefont {B.~A.}\ \bibnamefont
  {Kalinikos}}, \bibinfo {author} {\bibfnamefont {A.~B.}\ \bibnamefont
  {Ustinov}}, \ and\ \bibinfo {author} {\bibfnamefont {A.}~\bibnamefont
  {Stashkevich}},\ }\href {\doibase 10.1088/1742-6596/769/1/012071} {\bibfield
  {journal} {\bibinfo  {journal} {J. Phys. Conf. Ser.}\ }\textbf {\bibinfo
  {volume} {769}} (\bibinfo {year} {2016}),\
  10.1088/1742-6596/769/1/012071}\BibitemShut {NoStop}%
\bibitem [{\citenamefont {Boardman}\ and\ \citenamefont
  {Nikitov}(1988)}]{Boardman1988}%
  \BibitemOpen
  \bibfield  {author} {\bibinfo {author} {\bibfnamefont {A.~D.}\ \bibnamefont
  {Boardman}}\ and\ \bibinfo {author} {\bibfnamefont {S.}~\bibnamefont
  {Nikitov}},\ }\href@noop {} {\bibfield  {journal} {\bibinfo  {journal} {Phys.
  Rev. B}\ }\textbf {\bibinfo {volume} {38}},\ \bibinfo {pages} {11444}
  (\bibinfo {year} {1988})}\BibitemShut {NoStop}%
\bibitem [{\citenamefont {Lim}\ \emph {et~al.}(2018)\citenamefont {Lim},
  \citenamefont {Bang}, \citenamefont {Trossman}, \citenamefont {Amanov},\ and\
  \citenamefont {Ketterson}}]{Lim2018}%
  \BibitemOpen
  \bibfield  {author} {\bibinfo {author} {\bibfnamefont {J.}~\bibnamefont
  {Lim}}, \bibinfo {author} {\bibfnamefont {W.}~\bibnamefont {Bang}}, \bibinfo
  {author} {\bibfnamefont {J.}~\bibnamefont {Trossman}}, \bibinfo {author}
  {\bibfnamefont {D.}~\bibnamefont {Amanov}}, \ and\ \bibinfo {author}
  {\bibfnamefont {J.~B.}\ \bibnamefont {Ketterson}},\ }\href {\doibase
  10.1063/1.5007263} {\bibfield  {journal} {\bibinfo  {journal} {AIP Adv.}\
  }\textbf {\bibinfo {volume} {8}},\ \bibinfo {pages} {056018} (\bibinfo {year}
  {2018})}\BibitemShut {NoStop}%
\bibitem [{\citenamefont {Inglis}\ \emph {et~al.}(2019)\citenamefont {Inglis},
  \citenamefont {Tock},\ and\ \citenamefont {Gregg}}]{Inglis2019}%
  \BibitemOpen
  \bibfield  {author} {\bibinfo {author} {\bibfnamefont {A.}~\bibnamefont
  {Inglis}}, \bibinfo {author} {\bibfnamefont {C.~J.}\ \bibnamefont {Tock}}, \
  and\ \bibinfo {author} {\bibfnamefont {J.~F.}\ \bibnamefont {Gregg}},\ }\href
  {\doibase 10.1007/s42452-019-0500-x} {\bibfield  {journal} {\bibinfo
  {journal} {SN Appl. Sci.}\ }\textbf {\bibinfo {volume} {1}},\ \bibinfo
  {pages} {480} (\bibinfo {year} {2019})}\BibitemShut {NoStop}%
\bibitem [{\citenamefont {Prabhakar}\ and\ \citenamefont
  {Stancil}(2009)}]{Prabhakar2009}%
  \BibitemOpen
  \bibfield  {author} {\bibinfo {author} {\bibfnamefont {A.}~\bibnamefont
  {Prabhakar}}\ and\ \bibinfo {author} {\bibfnamefont {D.~D.}\ \bibnamefont
  {Stancil}},\ }\href {\doibase 10.1007/978-0-387-77865-5} {\emph {\bibinfo
  {title} {Spin Waves Theory Appl.}}}\ (\bibinfo  {publisher} {Springer US},\
  \bibinfo {address} {Boston, MA},\ \bibinfo {year} {2009})\BibitemShut
  {NoStop}%
\bibitem [{\citenamefont {Wu}\ \emph {et~al.}(2004)\citenamefont {Wu},
  \citenamefont {Kalinikos},\ and\ \citenamefont {Patton}}]{Wu2004}%
  \BibitemOpen
  \bibfield  {author} {\bibinfo {author} {\bibfnamefont {M.}~\bibnamefont
  {Wu}}, \bibinfo {author} {\bibfnamefont {B.~A.}\ \bibnamefont {Kalinikos}}, \
  and\ \bibinfo {author} {\bibfnamefont {C.~E.}\ \bibnamefont {Patton}},\
  }\href {\doibase 10.1103/PhysRevLett.93.157207} {\bibfield  {journal}
  {\bibinfo  {journal} {Phys. Rev. Lett.}\ }\textbf {\bibinfo {volume} {93}},\
  \bibinfo {pages} {157207} (\bibinfo {year} {2004})}\BibitemShut {NoStop}%
\bibitem [{\citenamefont {Khivintsev}\ \emph {et~al.}(2011)\citenamefont
  {Khivintsev}, \citenamefont {Marsh}, \citenamefont {Zagorodnii},
  \citenamefont {Harward}, \citenamefont {Lovejoy}, \citenamefont {Krivosik},
  \citenamefont {Camley},\ and\ \citenamefont {Celinski}}]{Khivintsev2011}%
  \BibitemOpen
  \bibfield  {author} {\bibinfo {author} {\bibfnamefont {Y.}~\bibnamefont
  {Khivintsev}}, \bibinfo {author} {\bibfnamefont {J.}~\bibnamefont {Marsh}},
  \bibinfo {author} {\bibfnamefont {V.}~\bibnamefont {Zagorodnii}}, \bibinfo
  {author} {\bibfnamefont {I.}~\bibnamefont {Harward}}, \bibinfo {author}
  {\bibfnamefont {J.}~\bibnamefont {Lovejoy}}, \bibinfo {author} {\bibfnamefont
  {P.}~\bibnamefont {Krivosik}}, \bibinfo {author} {\bibfnamefont {R.~E.}\
  \bibnamefont {Camley}}, \ and\ \bibinfo {author} {\bibfnamefont
  {Z.}~\bibnamefont {Celinski}},\ }\href {\doibase 10.1063/1.3541787}
  {\bibfield  {journal} {\bibinfo  {journal} {Appl. Phys. Lett.}\ }\textbf
  {\bibinfo {volume} {98}} (\bibinfo {year} {2011}),\
  10.1063/1.3541787}\BibitemShut {NoStop}%
\bibitem [{\citenamefont {Demidov}(1998)}]{Demidov1998}%
  \BibitemOpen
  \bibfield  {author} {\bibinfo {author} {\bibfnamefont {V.~E.}\ \bibnamefont
  {Demidov}},\ }\href {\doibase 10.1134/1.567807} {\bibfield  {journal}
  {\bibinfo  {journal} {J. Exp. Theor. Phys. Lett.}\ }\textbf {\bibinfo
  {volume} {68}},\ \bibinfo {pages} {869} (\bibinfo {year} {1998})}\BibitemShut
  {NoStop}%
\bibitem [{\citenamefont {Wu}\ \emph {et~al.}(2009)\citenamefont {Wu},
  \citenamefont {Hagerstrom}, \citenamefont {Eykholt}, \citenamefont
  {Kondrashov},\ and\ \citenamefont {Kalinikos}}]{Wu2009}%
  \BibitemOpen
  \bibfield  {author} {\bibinfo {author} {\bibfnamefont {M.}~\bibnamefont
  {Wu}}, \bibinfo {author} {\bibfnamefont {A.~M.}\ \bibnamefont {Hagerstrom}},
  \bibinfo {author} {\bibfnamefont {R.}~\bibnamefont {Eykholt}}, \bibinfo
  {author} {\bibfnamefont {A.}~\bibnamefont {Kondrashov}}, \ and\ \bibinfo
  {author} {\bibfnamefont {B.~A.}\ \bibnamefont {Kalinikos}},\ }\href {\doibase
  10.1103/PhysRevLett.102.237203} {\bibfield  {journal} {\bibinfo  {journal}
  {Phys. Rev. Lett.}\ }\textbf {\bibinfo {volume} {102}},\ \bibinfo {pages}
  {237203} (\bibinfo {year} {2009})}\BibitemShut {NoStop}%
\end{thebibliography}%

\end{document}